# Quantum Radar and Research Assessment
# (Preprint)

Gaspare Galati[1,2], Gabriele Pavan[1,2], Fred Daum[3]

[1]Department of Electronic Engineering, Tor Vergata University, Rome, - 00133, Italy

[2]CNIT (National Inter-University Consortium for Telecommunications),

RU of Rome Tor Vergata, Rome, 00133, Italy.

[3]Raytheon Company—System Engineering, Woburn, MA, 01801, USA.

*Abstract:* Quantum Radar was studied in many Nations for about fifteen years with the production of some hundred publications. In the post-2020 literature, it is shown that, due to the exceedingly low transmitted power, Quantum Radar cannot produce neither significant results nor real-world applications. Regrettably, most of the scientific community ignored this negative outcome: a fact worth of exam. A detailed study of such an *"assessment failure"* depicts the main shortcomings of the present situation, calling for a re-design of the research assessment at the international level, with proposals shown in the ending section of this paper.

## 1. Introduction

Since about 1990 quantum technologies are widely proposed (especially in the present "Quantum Year") as a key element for the world's development. These technologies include very different applications, some of which have demonstrated great success [1], [2] and [3]. In other cases, valuable results are to be assessed, as in Quantum Communications and Quantum Computation [4]. Finally, as discussed in the following, Quantum Radar (QR) techniques did never lead to any successful result despite some significant efforts. From a sociological point of view, such a particular failure can be analyzed in the frame of the *"falsification theory"* due to Karl R. Popper (1902-1994) [5], [6].

In the Histories by Thucydides (Thoukudídes, 5[th] century B.C., Athens) we read the still-true sentence: *"Most people don't bother to find out the truth but find it much easier to accept the first story they hear"*.

This paper aims to analyze a case of exploitation – by unadvertised (and sometimes, unscrupulous) operators – of these human weaknesses in a specific technical/scientific (and social) context, and to draw conclusions about the needed improvements of the worldwide research assessment.

## 2. The Quantum Radar Story

Nature and operation of Radar, both conventional (i.e. classical) and quantum, are very briefly described in [7] and [8] for the interested reader. The basic concepts related to Radar and to optimal reception are resumed in Section 2.6 of [8] and in [9], [10]. The maximum Range of a Radar depends on the echo energy, which is proportional to the transmitted energy. The energy of a single microwave photon – at the exemplary wavelength $\lambda = 3\ cm$, ($f_0 = 10\ GHz$, X-band) – is of the very low order of $6 \cdot 10^{-24}$ J.

For a 200 MHz band the corresponding power is of the order of $10^{-15}\ W$, calling for an exceedingly large correlation time $T$, not compatible with target's motion.

The results shown in [7] and [8] are extended here to the most used radar frequency bands. **Figure 1** shows the limited Range of a Quantum Radar (QR) versus the illumination time $T$, as compared to a Classical Radar (CR). The pertaining computations use the well-known *radar equation*, [9].

One may wonder which value for the observation time $T$ of the abscissa applies to practical applications. The answer is easily found considering the radial speed $v_R$ of the target of interest, in formulas: $T \leq \frac{1}{2B} \cdot \frac{c}{v_R}$ where $B$ is the operating bandwidth and $c$ the speed of light. Typical values of observation time, as shown in **Table 1**, justify the choice of the abscissa span in **Figure 1**.

A former analysis of quantum effects on radar is due to two DARPA (Defense Advanced Research Projects Agency) Scientists [11]. An early Quantum Illumination concept is found in two 2008 papers, [12] and [13], and in an USA patent (assignee: Lockheed Martin Corporation) [14]. The QR is a proposed microwave implementation of the Quantum Illumination (QI) concept, whose operation is resumed as follows. QI transmits quantum signal states of electromagnetic radiation that are entangled with quantum *ancillary* states that are kept at the transmitter, showing an entangled pair of photons, one stored in the transmitter and another, reflected from the target, sent to a measuring (correlation) device.

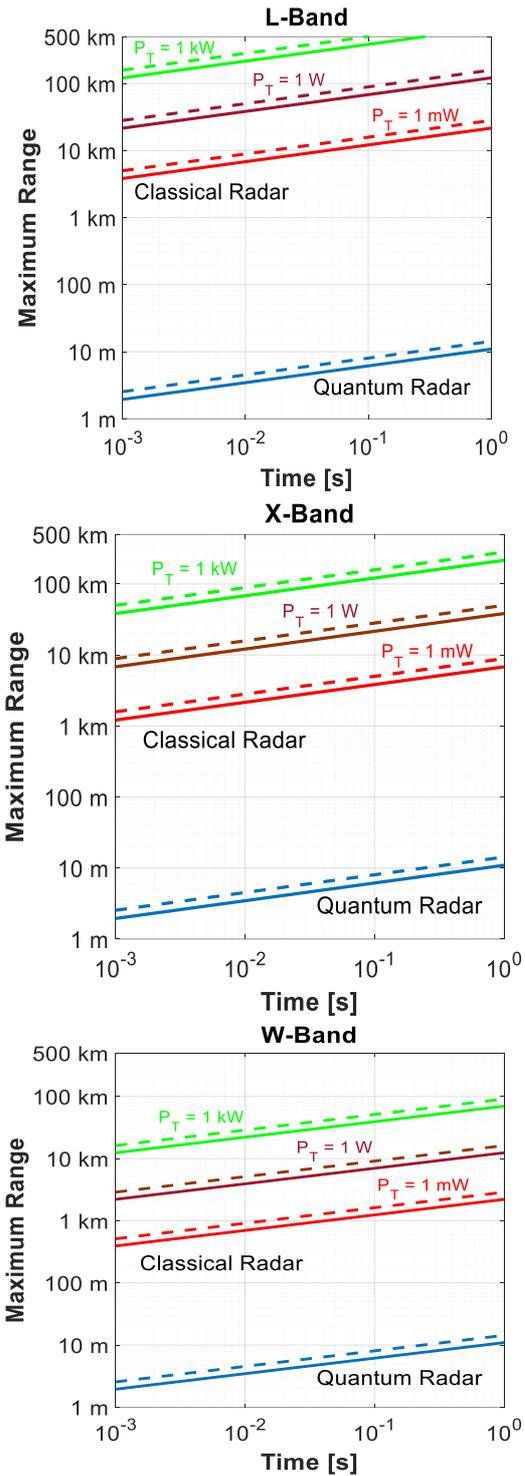

| Type of Target | $v_R$ (m/s) | T (s) B = 100 MHz | T (s) B = 1 GHz |
|---|---|---|---|
| Mach-3 aircraft | 1000 | 0.0015 | 0.00015 |
| Mach-1 aircraft | 300 | 0.005 | 0.0005 |
| Fast ship 30 kt | 15 | 0.1 | 0.01 |
| Slow ship 10 kt | 5 | 0.3 | 0.03 |

**Table 1.** – Maximum observation time for different targets

The history of QR research can be divided into four, partly overlapped, periods, as described in [8]. Concerning the *"Proposals phase"* (2008-2011), early writings in this period show generic proposals for QR. Their forerunner is the book *Quantum Radar* [15]. An interesting addition to the story of the QR is the, apparently not obvious, connection between three entities: (a) the Italy-based Leonardo Company, https://www.leonardo.com/en/home, (b) the author of [12], Seth Lloyd, and (c) Quantum technologies (in particular, QR). This connection is found after visiting Lloyd's official Curriculum Vitae (CV) on the MIT web site (accessed on 04/04/2025): https://meche.mit.edu/people/faculty/SLloyd@MIT.EDU. In the chronological part of Lloyd's CV we read:

| *Rank* | *Beginning* | *Ending* |
|---|---|---|
| Assistant Professor | December 1994 | June 1998 |
| <u>Finmeccanica Career Development Professorship</u> | <u>September 1996</u> | <u>Present</u> |
| Associate Professor (without tenure) | July 1998 | June 2001 |
| Associate Professor (with tenure) | June 2001 | June 2002 |
| Professor | June 2002 | Present |

Hence, Finmeccanica (now Leonardo Company), since 1996 is spending an alleged amount of the order of one million $ per year for Lloyd's academic chair. It should be interesting to investigate the industrial return of this investment, now close to its 30[th] year.

In the second phase described in [8], the *"Early papers production"* (2011-2017), no real data nor analyses support the claims of the effectiveness of a microwave QR and experimental data are practically absent as well as any quantitative evaluations of the maximum QR Range. This is a clear case of publication bias, i.e. one of the many unfortunate biases affecting research [16].

A third phase, the *"Mass papers production"* (2018-2021), is characterized by a peak in the QR publications. A search for QR on IEEEXplore (the database of the Institute of Electrical and Electronic Engineers, which includes more than 6.1 million items) yielded the results shown in **Figure 2**. On QR there are 132 papers (starting from 2012) distributed on Journals (28), Magazines (20) and Conference Proceedings (84). Considering the papers not referenced in IEEEXplore, the total largely exceeds a few hundred publications. A significant result of the analysis is that the unique book [15] is *antecedent* to all the IEEE found papers and to most of the overall papers. The years 2020 and 2021 show the maximum interest concerning QR. This interest was rapidly decreasing in 2023 – 2025.

**Figure 1** Comparison between the Maximum Range for Continuous-Wave CR and for QR, at L, X, W-band ($f_0 = 1, 10, 95$ GHz) vs. the time-duration $T$. System noise temperature: 290 K (solid line) and 100 K (dashed line). The CR transmitted power is 1 mW, 1 W, 1 kW, with an antenna gain of $G = 30\ dB$. Target's Radar Cross Section is $1\ m^2$, the overall losses are $L = -4\ dB$, the bandwidth is the 10% of $f_0$, the minimum Signal-to-Noise Ratio (SNR) is 13.2 $dB$. The quantum advantage is assumed equal to the reciprocal of the average number of photons per mode (less than the unity). Ideal propagation, no horizon, no atmospheric attenuation, no interference.

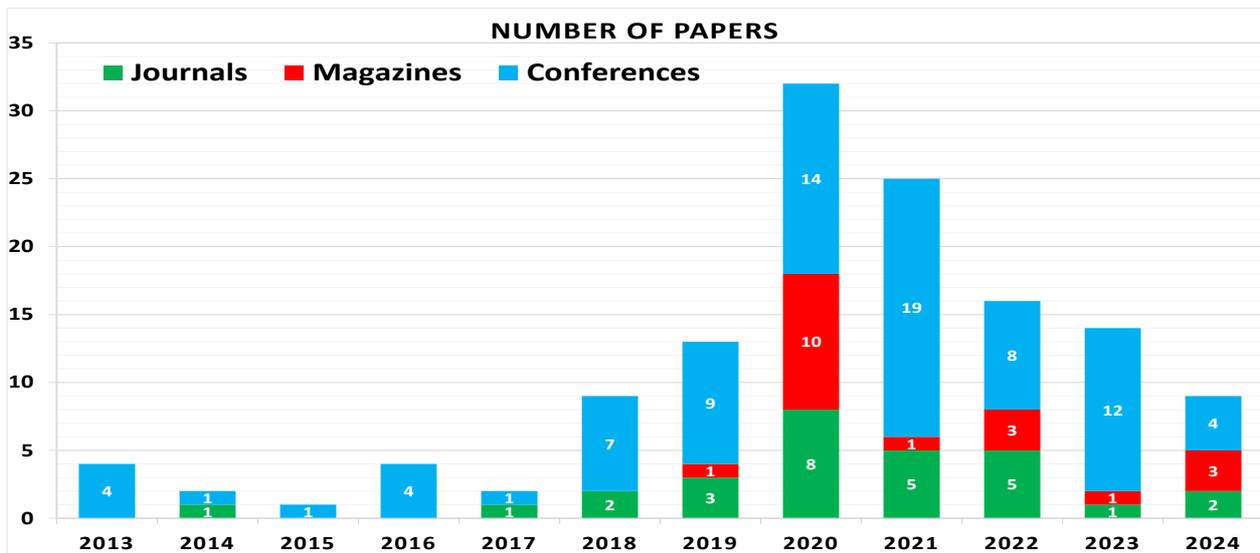

**Figure 2.** Papers on QR year by year referenced in IEEEXplore (https://ieeexplore.ieee.org/ Xplore/home.jsp).

The *"Disillusionment and endurance"* (2021-2025) phase shows some early disillusionment (it is recognized that the laws of physics make it impossible to implement a Long-Range QR and Endurance, as in [17] and [18]). The level of announcement remains high in some papers, for example in [19] we read *"... a potential hybridization between cognitive radar and quantum radar, a CQR. We feel that the quantum remote sensing will be deployed ... We are in the stage of proving the concept and realizing even more powerful test beds"*, and in [20] and [21] *"Quantum radar systems ... extend their capabilities ... encompassing the detection and identification of RF stealth platforms ..."*.

Even in 2025, some QR noticing remains on the Internet (see for instance, accessed on 24/04/2025: *Quantum Radar: the secret startups don't want you to know*, YouTube, Franks World of AI, 19 April 2025, https://www.youtube.com/shorts/9-QWanHsKlA and *The Game-Changing Potential of Quantum Radar- Explore the revolutionary technology of quantum radar, its impact on industries and society,* https://www.youtube.com/watch?v=VriVJkWjH9Q).

Summing up, the title of [22]: *The short, strange life of quantum radar*, tells us, once again, that while it is generally difficult to find the correct solution for a given problem, it may be even harder to find the correct problem for a given solution.

Noticeably, some QR researchers are connected each other in *"enduring groups"* operating according to both sides of the coin: (i) continuing publications on QR (and getting research funds), and (ii) preventing other groups to publish contrasting results and considerations. The most used tools by these *groups* are: (i) to control the acceptance of favorable submitted papers (including the suppression of the effects of negative reviews) and (ii) to control the rejection of submitted papers when they should impair some QR proposals. More details and exemplary cases are presented in [23].

More generally, social networks are a favorable seat for bloggers and marketing professionals who push false narratives about what quantum technologies may do or when they will arrive. In LinkedIn many examples were found by a search (on 16/04/2025) on #QuantumRadar (or #quantumradar or #quantumradarsystems).

For an updated and unbiased review on QR, one may see [24] where we read: *"...it is extremely unlikely to foresee any near-term development of a long-range quantum radar. Besides the distance limitations, there are the challenges associated with the actual ranging capabilities and detection times"*. For military applications of quantum technologies, we recommend the RAND report [25] were we read *"... quantum radar proposes to use radio-frequency entangled photons for long-distance ISR. However, the Defense Science Board has concluded that quantum radar will not provide upgraded capability to the Department of Defense (DoD)"*.

A presentation by Dr. John Burke (former Principal Director of quantum technologies at the Pentagon) was given on the same topic at the 12$^{th}$ Annual Meeting of the National Academy of Invention, Washington DC on 25 September 2023. This video https://www.youtube.com/watch?v=5-uJzSP-ULE (accessed on 10/04/2025 – the quantum radar is treated at 20' 35'') is accompanied by **Figure 3**, where the QR is in the *"impractical"* region.

Most QR papers are oriented to quantum physics and to related technological aspects, while very few contributions consider system and operational concepts. More details on QR literature are in [7], [8] and [26], [27], [28]. Still in 2025, confusion is present between the optical LADAR (or LIDAR) and the microwave (radiofrequency) Radar.

The recent paper [4] deserves comments. The paper has 9 authors and 454 References, 91 (i.e. 20%) out of them being by at least one author of [4], and 30 (6.7 %) by the first author (Hanzo). The *"self-referencing"* count averaged over the authors is 12%. Concerning the Quantum Radar part of [4] and its References from [240] to [271], it is interesting to notice that 23% of them are co/authored by the co-author of [4] Ivan B. Djordjevic. On page 51 Djordjevic declares 590 own papers (most probably written in 1999-2024, i.e., a production of about *two papers per month*) plus 11 books and 13 books-chapters. Note that his experience is optical: nothing on radar.

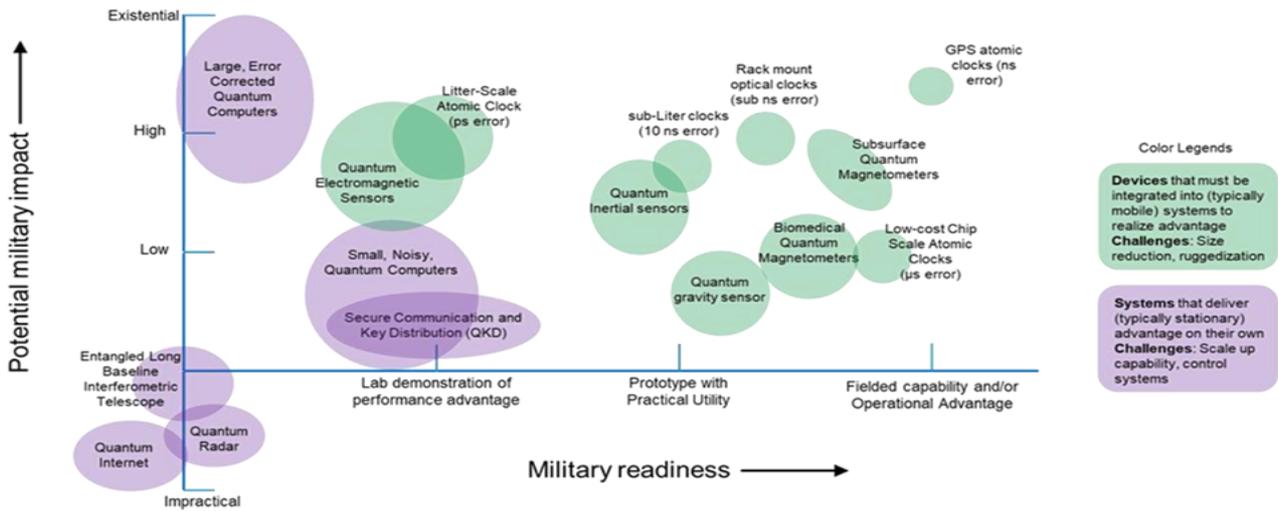

**Figure 3.** Military Technologies: Potential impact vs. readiness (by John Burke).

Summing up, in [4], whose overall scientific quality is not analyzed here, we found the following symptoms, whose general context is treated in the following.
- The improper selection of References neglecting, inter alia, all QR criticizing papers.
- The Radar-Lidar (wanted?) confusion.
- The over-signature of papers.
- The *"firm of signatures"* effect, i.e. a purposely addition of co-authors.

The overview and *scientometric* paper [29] does not consider Quantum Radar at all. The same applies to [30]. This is not surprising as (i) it was a niche research area (about ten papers per year versus the ten thousand papers per year in the overall quantum technologies), and (ii) at the time of writing of [29], half of 2021, the QR was going to start its decay. Anyway, the QR research, *three orders of magnitude* smaller than the overall quantum, is interesting as a kind of *"thermometer"* of disease in the national and international research context, a topic well described in [16].

The *"disillusionment and enduring"* phase of QR research, with a widespread opinion about the impossibility of a Long-Range QR, includes the attempt to propose an impractical, costly and useless *"short-range"* biomedical QR [18].

Somewhat similarly, the much larger community working on Quantum Computer (QC), after many claims of *"supremacy"* on classical computers (and on supercomputers) in the recent years, proposes particular systems called NISQ (Noisy Intermediate Scale Quantum), which are *"officially"* prone to decoherence and to quantum errors, i.e. the main reasons for the lack of effectiveness of any Quantum Computer. About them, see for instance:
https://en.wikipedia.org/wiki/Noisy_intermediate-scale_quantum_era (accessed on 12/05/2025). The related claim by the proposers refers to the solution of *"specific problems not solvable by traditional computers"* [31]. An example is the Advantage2$^{TM}$ by D-Wave, www.dwavequantum.com (accessed on 28-05-2025) whose task is *"solving 3D lattice-spin glasses of tunable precision"*: nothing to do with any realistic task for digital computers.

## 3. More Stories with Wrong Ways

### 3.1 The "Cold" Nuclear Fusion

QR is an example of *"wrong way"* in research and development. Its story (another example of worldwide unrealized promise) is someway similar to that of Cold Nuclear Fusion. The alleged energy production by nuclear fusion at (or near) room temperature, or *"cold fusion"* (CF), later renamed LENR (Low Energy Nuclear Reaction, https://lenr-canr.org/ , accessed on 07/04/2025) dates back to the announcement (presented as a route to unlimited and clean energy) on March 1989 by M. Fleischmann and S. Pons, electrochemists at the University of Utah. The early alleged results of their experiment raised hopes of a cheap and abundant source of energy but were not confirmed by independent trials. However, on August 1989, the State of Utah invested 4.5 million of US $ to create the National Cold Fusion Institute (closed on June 1991, after it ran out of funds). In September 1990, the National Cold Fusion Institute listed 92 groups of researchers worldwide that had reported evidence of excess heat, but refused to provide any details and any evidence of their results arguing that they could endanger their patents. In spite of those researcher's behavior, well described in the section *"The pervasive dishonesty of modern science"* of [16], some support was found in the general press such as in [32]. Papers related to LENR and related topics are still present in the recent literature [33], [34], and an industrial application is proposed with the name E-Cat by Andrea Rossi,

https://en.wikipedia.org/wiki/Andrea_Rossi_(entrepreneur), https://e-catworld.com/, and https://e-catworld.com/2025/05/07/rossi-reports-tests-completed-partner-fully-convinced-deliveries-to-start-in-2025/ (all accessed on 28/05/2025).

In 2015, Google Company *quietly* organized chemists and physicists from several universities and opened its own labs to cold fusion. However, in spite of the fact that the Google's 10 million US $ project followed three different approaches, no experiment showed evidence of nuclear reactions nor of production of additional energy.

*3.2 Quantum Technologies: Promises and New Age Claims*

The so-called *"second quantum revolution"*, aimed to exploit the possibility of addressing and controlling single quantum entities (and widely using the keywords *"quantum superimposition"* and *"entanglement"*) is forty years old, as its origin may be dated back to the publication of the cryptographic protocol BB84 based on pairs of photons [35].

A related *"quantum fever"* is well present in the book [36], starting with the title of its first chapter: *"Mighty Power: How a Theory of the Microcosm Changed Our World"*, as well as in New Age and pseudo-scientific works such as [37], [38] and the unfruitful attempts to find a quantum operation in the human brain in [39-45].

Researchers and companies like Google's Quantum AI Lab and Nirvanic are exploring whether quantum computing can help unravel the mysteries of consciousness. Nirvanic Consciousness Technologies aims to merge quantum computing, artificial intelligence (AI) and quantum theories (https://www.nirvanic.ai/, accessed on 07/04/2025). Hartmut Neven, a physicist and computational neuroscientist leading Google's Quantum AI Lab, proposes leveraging quantum computers potentially expanding our understanding of how the mind interacts with the physical world. How to connect the human brain to a quantum computer (and to which one) is not explained.

For a recent example on quantum computing, please see [85-88] . Another amusing example is shown on YouTube (accessed on 31/05/2025)
(https://www.youtube.com/watch?v=UtDllX_MTbw ).

*3.3 Solutions for Problems or Vice-versa?*

It may be interesting to view the second quantum revolution [36] in the historical frame when the main technical and scientific developments – mostly, *"unpromised"* – have been conceived and implemented in order to solve real world problems. They include microelectronics, laser, LCD and LED monitors rubidium/cesium clocks, GPS and more from the *"first quantum revolution"*. Many recent technical-scientific revolutions were not *"promised"* (i.e. they were not announced by anybody) and their results are around us or in our pocket.

On the other side, some research efforts in quantum-related developments belong to the reverted paradigm of *"a solution calling for a problem"*. The QR is a good example of *"promised solution calling for an alleged problem"*, the fear of detection and location of stealth targets as analyzed above. The Quantum Computer is treated in the ensuing paragraph.

*3.4 Quantum Computer Efforts*

A possible example of *"solution calling for a problem"* is Quantum Computer (QC), as discussed in [46-49]. This feature is clearly (and indirectly) confirmed by one of the main private QC financers, that is, Google (the same company which funded LERN, paragraph 3.1). The Google-GESDA-XPRIZE (https://www.xprize.org/prizes/qc-apps, accessed on 12/05/2025) is a five million dollars global competition *for defining quantum computer applications to solve real world problems*: difficult not to see in it the paradigm of a solution calling for a problem.

A recent proposal in Quantum Computation is that of *"certified randomness"* i.e. the generation of sequences of random numbers, see [50]. It is true that, as stated in [50], *"Certified randomness has many applications but is impossible to achieve solely by classical computation"* because a computer may only generate pseudo-random numbers, but it is also true (see for instance https://en.wikipedia.org/wiki/Noise_generator, https://en.wikipedia.org/wiki/Noise_generator, accessed on 12/05/2025) that macroscopic phenomena such as the *thermal noise* from a Zener diode, or a vacuum tube, are *"as random as"* some proposed sophisticated and costly quantum sources.

Standard, commercial *noise sources* produce certifiable random numbers after a simple Analog-to-Digital conversion. Details are amply available in the open literature, an example being: https://tubedata.milbert.com/sheets/046/k/K81A.pdf.

Quantum Computer is over forty years old [51-53] and still not a single, general, usable set is on sale. In Sectio 2 *"On the advantage of quantum computing"* of [54], we read *"One big problem is the growth of the error rate with the number of entangled qubits: the most recent attempt at factorizing the number 35 on the IBM Q-System-One failed because of error accumulation (Amico et al., 2019, Experimental study of Shor's factoring algorithm using the IBM Q Experience. Phys. Rev. A 100, 012305. doi:10.1103/physreva.100.012305)"* (accessed on 28/05/2025)

In fact, after the theoretically unwarranted fanfare about the quantum algorithm for Shor's factorization in 1994 and after some attempts with optimization methods, no much more algorithms with a similar potential were found.

Research & Markets predicted in May 2021 that the global quantum technology market would even reach $31.57 Billion by 2026, including $14.25 Billion for quantum computing. Considering the investors' FOMO (Fear of Missing Out), the total quantum technologies start-up funding was slightly below $4 Billion worldwide.

*3.5 Investments in Quantum Technologies*

In spite of some decay of the funding and of the number of patents and of spin-offs, as shown in [7], [8], quantum fever continue to be well alive. The quantum-related investments arrived at significant values on both sides of the Atlantic. In Europe, the Quantum Flagship initiative is funded for 1 B€ (one billion Euro) in the period 2018-2028 [55]. **Table 2** resumes the total private investment by region ($ million).

|      | Americas | EMEA | APAC | Total |
|------|----------|------|------|-------|
| **2022** | 1369 | 762 | 260 | 2391 |
| **2023** | 240 | 781 | 217 | 1238 |

**Table 2.** Total *private* investment by region ($ million). EMEA: Europe, Middle East, and Africa; APAC: Asia-Pacific. Source: The Quantum Insider, updated at the end of December 2023, [55].

In the USA there is a competitive effort by Google, Microsoft and IBM accompanied by a large effort for the worldwide announcement – most in the Internet – of some alleged results.

According to [56] the investments on quantum technologies, evaluated on a five-year basis, are geographically divided as shown in **Table 3** (in billion Dollars, B$). From **Table 3** the overall global expense is of the order of over five billion Euro (B€) per year. Clearly, the suggestions to invest (which show more and more in the Internet) call for some care. According to [57] no one of the quantum computing companies or start-ups earns revenue from their products – be it hardware or software – unless we count D-Wave Systems' quantum *annealers*, which are not computers. In [58] some interesting proposals are described to mitigate the damaging effects of the current situation. Reference [59], from which we quote *"With this manifesto we wish to prevent quantum technology from running into fiascos of implementation at the interface of science and society"*, enhances the ethical aspects and proposes these key criteria for quantum technologies: Comprehensible, Specific, Open, Accessible, Responsible, Culturally embedded, Meaningful.

|  | Private (B$) | Public (B$) | Total (B$) |
|---|---|---|---|
| **USA** | 4.6 | 6.4 | 11.0 |
| **EU** | 6.6 | 2.0 | 8.6 |
| **China** | 2.3 | 1.0 | 3.3 |
| **Canada** | 0.8 | 1.3 | 2.1 |
| **United Kingdom** | 1.0 | 0.6 | 1.6 |

**Table 3.** Five-year investments (Billion $) in quantum technologies. From [58].

Summing up, while the QR situation is clearly assessed today, the *"three orders of magnitude larger"* Quantum Computer (QC) shows a much more complex status. In [46-48] it is explained why an effective, usable QC cannot be built. On the other hand, the big private and public investments and the production of some thousand papers per year put the QC in the group of initiatives *"too big to fail"*, similar to the (even bigger) Controlled Nuclear Fusion. Both initiatives show a time scale of about forty years as of today, i.e. longer than the one of the *much smaller* QR.

*3.6 Ethics and Freedom of Research*

Important ethical problems arise when announcing some (real or alleged) research results. An excerpt by R. Feynman from [60] reads: *"For example, I was a little surprised when I was talking to a friend who was going to go on the radio. He does work on cosmology and astronomy, and he wondered how he would explain what the applications of this work were. "Well," I said, "there aren't any." He said, "Yes, but then we won't get support for more research of this kind." I think that is somewhat dishonest. If you are representing yourself as a scientist, then you should explain to the layman what you are doing – and if they don't want to support you under those circumstances, then that's their decision"*.

In the video by Sabine Hossenfelder *"Quantum Hype Goes Crazy, But Why?"* [61] (240,000 views; accessed on 13.01.2025), it is interesting to notice the debate between Scott Aaronson (Chair of Computer Science at the University of Texas at Austin) and Edward Fahri from MIT (a physicist working on quantum computation at Google). According to Aaronson, for a scientist *"it is not enough not to say anything false"* and one of the comments on the video deserves a quotation: "It is always possible to write a paper in a way that obscures its weaknesses or makes them difficult to uncover. Authors can hide the flaws within complex explanations, and when reviewers request clarification, they can respond with even more sophisticated arguments … strategies like this are, in my experience, a common approach in academic writing".

A general result is the urgent need to restore what Feynman [60] calls the *culture of doubt* necessary for science to operate a self-correcting, truth-seeking process, as well as to counteract *"the corruption of real science"* as analysed in [16].

**4. Conclusions, Recommendations, and "Impossible" Proposals**

*4.1 Avoiding Further (and Bigger) Incidents*

The above analysis of the QR research shall lead to general considerations, teachings and proposals.

With the exponential increase of technical and scientific information, it is more and more difficult to separate meaningful and promising research plans, as well as ongoing research, from the *wrong-way* ones. From both social and ethical points of view, discovering and cutting away the *dry branches* in research and development is an urgent need.

Experience shows that *wrong-way* plans may be promoted by groups of researchers in *a bubble* keeping outside (ignoring or denying) any documented criticism, or even fighting it with all means, including the anonymous review of papers [16]. In [23] one may read that the use of these dual tools involves a celebrated professional Society, the IEEE, in two ways:

(i) Improper rejection of off-stream but correct manuscripts by MDPI Sensors (December-2023) and by IEEE-Access (April 2024).

(ii) Improper denial of the effect of correct peer reviews by IEEE Transactions on Radar Systems (January 2025).

Based on our recognized radar expertise, we have analyzed in detail the particular and *"financially small"* case of QR research, clearly lacking for any reasonable assessment. Being assessment, evaluation and funding critical elements, this single *"failure"* – as shown in the QR *saga* [7], [8] – should suffice to reconsider the whole research assessment process.

A safe and conservative assessment is routinely used in Air Traffic Control and Management regulations, by the International Standards and Recommended Practices (SARPS) contained in the ICAO's Technical Annexes to the Convention on International Civil Aviation. The ICAO Annex 13, *Aircraft Accident and Incident Investigation*, clearly states its aim: *"The sole objective of the investigation of an accident or incident shall be the prevention of accidents and incidents."*

In this conservative view, the fact that the scope of this work, of course, does not include the assessment of all *Big Science* projects does not prevent to draw general conclusions.

A well-known *Big Science* project is Controlled Nuclear Fusion. A rough estimate of the overall global expense (in billion Euros/Dollars) puts the Controlled Nuclear Fusion in the top range. In fact, considering only the largest international project, i.e. ITER (https://www.iter.org), the initial estimate amounts to 6 Billion € over ten years. However, after the 5 B€ overrun on July 2024, this estimate quickly gowned to 18 - 22 Billion € (other estimates arrive to 45 - 65 B€). Below this huge project, in the middle range of costs we find the Quantum Technologies (computation, sensing and communication), while QR is in the lowest range.

For Quantum Computer and Controlled Nuclear Fusion, the time horizon to arrive at valuable, practical and significant results continuously shift ahead putting out of control the cost/effectiveness ratios and the fairness due to the taxpayers.

Moreover, researchers and proposers of (real or prospective) unsuccessful (or *"never concluding"*) activities sometimes misuse the widely accepted paradigm of *"freedom of research"*, [62], [63].

Hence, and sad to say, avoiding the improper usage of *"freedom of research"* - when applied to *"non-research"* - is a first item to be considered when trying to improve the present, unsatisfactory situation.

*4.2 Discriminating Research from "non-Research"*

The *"mission impossible"* of separating – in a publicly recognized way – research from *non-research* is mandatory for a correct use of the money from the taxpayer and/or from the investor. For the definitions of *research* and of *non-research*, please see (accessed on 10/04/2025) https://erau.edu/research/irb/definitions-and-categories-of-research and [16].

Shortly stated, research shall develop (or contribute to) generalizable knowledge and be, first of all, *"falsifiable"* according to Karl Popper. For example, the often found [64] claim that *"consciousness is not an epiphenomenon of the brain activity and continues after death"* is not falsifiable (and not demonstrable). Second, a research shall not be fully based on grounds that have been assessed to be false (e.g. the possibility of building a long Range QR [7]), nor on indemonstrable thoughts and personal feelings, such as in [37] and [65].

A vast literature exist on the Assessment and Evaluation of Research, including [66] which contains motivated critics to the widely used bibliometric indexes, as well as [67] and [68]. In [66] problems such as misuse and misinterpretation of the impact factor as well as bibliometric manipulations (*ranking boosting*, *dummy affiliations*, and *intellectual fraud*) are treated.

Leaving the present situation "as is" will likely exacerbate the problems described in [69] by B. G. Charlton from the University of Buckingham (https://scispace.com/authors/bruce-g-charlton-360e0gt7rc, accessed on 10/04/2025). Sentences from [69] include the following *"Researchers are no longer trying to seek and speak the truth. Scientists no longer believe in the truth. ... Hence, the vast structures of personnel and resources that constitute modern science are not real science but merely a professional research bureaucracy"*. The huge, quickly increasing count of papers does not help the analysis of real and significant achievements, needed to discriminate *Research* from *non-Research*. On the other hand, the approval of manuscripts submitted for publication generates fundamental data for assessments. Therefore, review procedures deserve attention. The mention by Charlton of the *"predictive and fast"* peer review and of the *"retrospective and slow"* peer usage may suggest a *"mission impossible"* solution to the review problem, with focus on the users and based on dedicated teams.

*4.3 Reviews and Bibliometric Indexes*

To accept/correct/reject the submitted manuscripts most Journals use the *"peer review"* often preceded by an Editor review aiming to check whether the manuscript fits the Journal.

Journals use various types of peer review, among them: (i) double blind (Authors and Reviewers remain anonymous), (ii) blind Reviewers with known Authors, (iii) open Reviewers who publish and sign their reviews, and more. Examples are found in https://libraryguides.mcgill.ca/journalpublishing/typesofreview (accessed on 10/04/2025). In most cases, the Reviewers do not receive rewards for their time and effort. Of course, a voluntary review does not always guarantee the adequate level of quality: during their career, many scholars have seen a wide range of reviews, from professional and helpful to *"quick and dirty"*.

Summing up, the present peer review system shows significant problems, see for instance [70]. It does not always block the publication of wrong or fabricated results such as, for example, in https://theconversation.com/problematic-paper-screener-trawling-for-fraud-in-the-scientific-literature-246317 - (accessed on 09/04/2025) and has been criticized [71] as *"slow, expensive, easily abused, prone to bias, unable to detect fraud"*.

Bibliometric indexes, used for enrolling researchers and for the advancements of careers, have an important effect on the academic system. Unfortunately, these indexes are prone to fraud and misuse, see for example the comments to [4] in Section 3. Their usage for rankings and for advancements has produced three damaging effects. The former is the multiplication of papers by splitting a single, significant content into two or more (less significant) papers. The second one is the multiplication of signatures, i.e. the addition of more signatures even when the paper was produced by a very few persons (or by a single author). The third one is a purposely-managed list of references, aimed to increase the citations count of the authors or of some selected colleagues.

According to the National Science Foundation (USA), https://ncses.nsf.gov/, (accessed on 10/04/2025) in 2023 the number of papers worldwide (including reviews, conference proceedings, and short surveys) in the areas of Science and Engineering (S&E) reached 5.14 million (about 1400 daily) with an exponential growth of about 5.6% per year. These figures can be compared with the overall production of academic papers (not only S&E) arriving to the amount of 90 million in the Web of Science, a database of scholarly articles from 22,000 peer-reviewed journals worldwide. The huge data bases of publications and the various related information have triggered a large number of statistical and economical studies, see for instance [72], [73] and [74].

The scientific community seems to accept the multiplications of papers and of signatures, despite exemplary cases such as the one of Rafael Luque Alvarez de Sotomajor, professor of chemistry at the Universidad de Cordoba, showing in scholar.google.ca/citations (accessed on 10/04/2025) a list of his publications. He declares to have published 1263 papers in 25 years, 58 out of them from January 2023 to April 2023, i.e. *over three papers per week*. More on Rafael is found in [75].

The present way of research assessment and funding is critical, so much that [76] and [77] suggest that it could be improved by adding a suited degree of random choices (a kind of *"lottery"*) to the imperfect judgements by experts. The limits of the Performance-Based Research Evaluation are evidentiated in [78].

**5. (Im)Possible Solutions**

*5.1 The overproduction of papers*

The increasing number of published papers and the declining average quality indicate that the review of manuscripts submitted for publication has to be definitely improved. More selective acceptance criteria are needed, without negating the freedom of publication as a part of the *"free speech"* rights.

The point of balance could be the quality control of each submitted scientific/technical public document by a permanent team of experts in its specific area (START-Scientific and Technical Advisory Review Team). A START evaluates each document according to public criteria such as (i) Originality and Novelty, (ii) Clarity and Completeness, (iii) Verifiability of results, (iv) Scientific/Technical value, (v) Relevance and Interest for the Scientific and the Users Community.

For each criterion, the START will set a score from 1: Unacceptable, to 5: Excellent.

A START will define a work *"worth of publication"* when: (a) all scores are greater than one, and (b) the sum of scores is greater than a given threshold (say, thirteen or fourteen). To limit the variability among STARTs, the scores are normalized to a Gaussian distribution whose mean and variance are defined a priori on the basis of the desired *"worth of publication"* rate and of the *"excellent"* rate.

The scores and their motivation will be available to the involved parties and will be printed on the cover of the published paper, thus granting:

a) to the Editors and Authors, the freedom of publishing (in principle, unacceptable documents may be published).
b) To the Readers (the final users), an independent information for a correct choice.

The constitutionally granted freedom of communication will remain available to everyone to submit his work as well as to any Editor (or Conference Chair) to accept (or reject) it.

Within this frame:

(i) Any user, before buying (or accessing to) a publication, can choose (or discard) it in an informed way.
(ii) The related information is by a thrusted and independent party.

*5.2 The overproduction of signatures*

The number of signatures of scientific papers is quickly increasing. For instance, in [79], studying the relation between number of authors and impact of a publication, the variation over time is analyzed in two close triennia: 2004-2006 and 2008-2010. The average number of authors per publication in the two triennia is presented in Table 7 of [79], showing an average increase of 10.0%. In more detail 9.6% for Engineering, 8.4 % for Physics, 11.5% for Earth and Space Sciences: a noticeable and quick increase, indeed.

An obvious, theoretically simple solution to the overproduction of the paper's signatures (which enter into the bibliometric systems) should be twofold: (i) to weight the signatures and (ii) to limit the signatures per unit time. Item (i) means that each co-author of a paper shall declare the own percentage contribution, as agreed within the author's team. Each contribution shall show as an integer percentage *"weight"* from 1 to 99. The number of publications by each scholar shall be the result of a *"weighted"* sum.

Item (ii) consists in limiting each author's contribution to the assessment (based on the aforementioned scores) to a reasonable upper limit per year, selecting his/her best papers up to the amount of, say, ten.

These *"Author-side"* regulations, today *"impossible"* for lack of a proper Authority (the equivalent of ICAO for Research does not exist), should be complemented by according *"Editor-side"* ones, not discussed here.

*5.3 Research Assessment Teams and International Agencies*

In the frame of these *impossible proposals*, all research groups should undergo effective, timely and open assessments for all phases of their research projects (proposal, planning, development, documentation and delivery to the users).

The (public or private) funding body shall take into account of open and independent assessments by the relevant STARTs, enriched by additional criteria tailored to the specific type of research.

The START members shall be independent, competent, blackmail-free, full-time experts, covering all the active and potential research areas and operating like a Coroner's jury. These scholars and researchers dedicating their full-time to assessments shall be equated to civil servants for the whole duration of their charge (say, five years after their retirement) and will be adequately paid for that.

How to enroll them is an open question justifying the adjective *"impossible"* in the title of Section 5: it should be relatively easy in an industrial and research context well controlled by a *monocratic* State, as in China, Russia or Iran, but more difficult in other contexts.

The research assessments, which include both *"predictive"* and *"retrospective"* aspects, will evolve with time, along with the development of each research under exam. The time span should at least follow the duration of the *Gartner cycle*, as shown in [8]. Of course, this duration varies with the burden – and cost – of the research project. Relatively small projects, such as QR, show a duration of the order of five to ten years (**Figure 2**) which could be correlated with a cost of the order of one million € (or $). The more costly Nuclear Cold Fusion showed a duration of the order of ten to twenty years. The Controlled Nuclear Fusion, not yet arrived to the (likely) disillusionment, shows a cost of Billion € and a duration of over half a century: the - politically hard - recognition of a failure of such a big project tends to be delayed as much as possible (due to the *"too big to fail"* concept). A similar situation, in our opinion, will apply to Quantum Computer, whose concept, anyway, is over forty years old [80].

The dream of an International Research Evaluation does not appear really impossible, as the last decades have seen attempts to establish international bodies for research assessment. One of them is CoARA - Coalition for Advancing Research Assessment, [81]. Another relevant initiative is the Declaration on Research Assessment (DORA), which recognize the need to improve the assessment of scholarly research beyond the bibliometric indexes, (https://sfdora.org/, accessed on 24/04/2025), see also [82], [83].